\documentclass[journal]{IEEEtran}
\makeatletter

\newcommand{\Rmnum}[1]{\expandafter\@slowromancap\romannumeral #1@}
\makeatother

\usepackage{mathrsfs}
\renewcommand{\IEEEQED}{\IEEEQEDopen}

\usepackage{ulem} 
\usepackage{float}
\usepackage{stfloats}
\usepackage{dsfont}
\usepackage{cite}
\usepackage{graphicx}
\usepackage{subfigure}
\ifCLASSINFOpdf

\else

\fi
\usepackage[cmex10]{amsmath}
\usepackage{float}
\usepackage{array}
\usepackage{algorithm}  
\usepackage{algorithmic}  
\usepackage{amsmath}
\usepackage{amsfonts}
\usepackage{amssymb}
\usepackage{ulem}
\usepackage{cancel}

\usepackage{xcolor}
\ifCLASSINFOpdf
\else
   \usepackage[dvips]{graphicx}
\fi
\usepackage{url}

\usepackage{graphicx}

\begin{document}

\title{{Hyper-Parameter Auto-Tuning for Sparse Bayesian Learning}}

\author{Dawei Gao, 
Qinghua Guo, 
Ming Jin, Guisheng Liao, and Yonina C. Eldar 
\thanks{Corresponding to Qinghua Guo (qguo@uow.edu.au).}
\thanks{Dawei Gao and Guisheng Liao are with the Hangzhou Institute of Technology, Xidian University, Hangzhou 311200, China and also with the National Laboratory of Radar Signal Processing, Xidian University, Xi'an 710071, China
(e-mail: \{gaodawei, liaogs\}@xidian.edu.cn).}
\thanks {Qinghua Guo is with the School of Electrical, Computer and Telecommunications Engineering, University of Wollongong, NSW 2522, Australia (e-mail: qguo@uow.edu.au).}
\thanks {Ming Jin is with the Faculty of Electrical Engineering and Computer Science, Ningbo University, Ningbo 315211, China (e-mail:jinming@nbu.edu.cn). }
\thanks{Yonina C. Eldar is with the Faculty of Math and CS, Weizmann Institute of
Science, Rehovot, 7610001, Israel (email: yonina.eldar@weizmann.ac.il).}
}

\markboth{}
{Shell \MakeLowercase{\textit{et al.}}: Bare Demo of IEEEtran.cls for IEEE Journals}
\maketitle

\begin{abstract}
Choosing the values of hyper-parameters in sparse Bayesian learning (SBL) can significantly impact performance. However, the hyper-parameters are normally tuned manually, which is often a difficult task.  
Most recently, effective automatic hyper-parameter tuning was achieved by using an empirical auto-tuner. 
In this work, we address the issue of hyper-parameter auto-tuning using neural network (NN)-based learning. Inspired by the empirical auto-tuner, we design and learn a NN-based auto-tuner, and show that considerable improvement in convergence rate and recovery performance can be achieved.
\end{abstract}

\begin{IEEEkeywords}
 Sparse Bayesian learning, hyper-parameter, neural networks.
\end{IEEEkeywords}

\IEEEpeerreviewmaketitle

\section{Introduction}

\IEEEPARstart{W}{e} consider the use of sparse Bayesian learning (SBL) to recover a length-$N$ sparse vector $\mathbf{x}$
from measurements
\begin{equation}\label{sysm}
    \mathbf{y} = \mathbf{Ax} + \boldsymbol{\zeta}, 
\end{equation}
where $\mathbf{y}$ is a length-$M$ measurement vector, $\mathbf{A}$ is an $M \times N$ measurement matrix, $\boldsymbol{\zeta}$ is a Gaussian noise vector with mean zero and covariance matrix $\beta^{-1}\mathbf{I}$, and $\beta$ is
the noise precision.{This problem finds numerous
applications in various areas of signal processing, statistics and
computer science \cite{eldar_2015,eldar2012compressed,duarte2011structured,liu2019sparse,candes2006near,wen2017efficient}. }

In SBL \cite{tipping2001sparse}, two-layer sparsity-promoting priors are used for $\mathbf{x}$, i.e.,
\begin{equation}
	\begin{aligned}
		p(\mathbf{x}| \boldsymbol{\gamma}) &=\prod_n p({x}_n| {\gamma}_n)=\prod_n \mathcal{N}(x_{n}|0, \gamma_{n}^{-1}),
	\end{aligned}
\end{equation}
\begin{equation}\label{xg}
	\begin{aligned}
		p(\boldsymbol{\gamma}) &=\prod_n p( {\gamma}_n)=\prod_n \text{Ga}(\gamma_{n} |\epsilon, \eta),
	\end{aligned}
\end{equation}
{where the precision vector $\boldsymbol{\gamma}=[\gamma_1,\gamma_2,\ldots,\gamma_N]^T$, and ${\text{Ga}({\gamma}_n|\epsilon, \eta)}$ is a Gamma distribution with shape parameter $\epsilon$
and rate parameter $\eta$} (which are called hyper-parameters). 
The precision vector $\boldsymbol{\gamma}$ is learned by maximizing the logarithm of the {{a posteriori}} probability $p(\boldsymbol{\gamma}|\mathbf{y})$, and an iterative re-estimation strategy leads to  the iterative SBL algorithm \cite{tipping2001sparse}. 

The values of the hyper-parameters $\epsilon$ and $\eta$ are normally chosen empirically, e.g., they are set to  small values ($10^{-4}$) to make the priors non-informative \cite{tipping2001sparse}. However, {their values} can have  significant impact on the performance of the SBL algorithm, and it is highly desirable that the {hyper-parameters are} tuned automatically, rather than being chosen manually (which is often a difficult task if not impossible).  

The impact of the shape parameter $\epsilon$ on the performance of SBL was investigated in \cite{luo2021unitary}, {where} the following empirical auto-tuner for the shape parameter is {proposed}
\begin{equation}\label{ep}
	\epsilon =\frac{1}{2}\sqrt{\log(\frac{1}{N}\sum_n{\gamma}_n)-\frac{1}{N}\sum_n\log{\gamma}_n}.
\end{equation}
{This} is incorporated to the iterative process of SBL algorithms to update the shape parameter automatically. It {was} shown that the empirical auto-tuner is very effective, and it works well for different sparsity rates,  measurement matrices and signal to noise ratios (SNRs). To address the issue of high complexity of the conventional SBL algorithm in \cite{tipping2001sparse}, where a matrix inverse is required in each iteration,  a low-complexity SBL algorithm called unitary approximate message passing (UAMP)-SBL was also proposed in \cite{luo2021unitary}. {The empirical auto-tuner \eqref{ep} works well for both the conventional SBL and UAMP-SBL \cite{luo2021unitary}.}

In this work, inspired by the empirical auto-tuner, we aim to find better hyper-parameter auto-tuners leveraging neural network (NN)-based learning. In particular, we find a NN architecture and train a NN-based auto-tuner by unfolding the iterative (UAMP-)SBL algorithm, and the NN-based auto-tuner is trained through back-propagation with tied parameters. Once the NN-based tuner is learned, it is incorporated to the iterative process of the (UAMP-)SBL algorithm. We show that better convergence rate or recovery performance can be achieved by using the learned NN-based auto-tuner.

The remainder of the letter is organized as follows. In Section II, we briefly review the conventional SBL and UAMP-SBL algorithms. In Section III, the design and training of the NN-based auto-tuner are elaborated. Simulation results are provided in Section IV, followed by conclusions in Section V. The notations used are as follows. Boldface lower-case and upper-case letters denote vectors and matrices, respectively. The superscripts 
$(\cdot)^T$ and $(\cdot)^H$ represent the transpose and conjugate transpose operations, respectively. 
We
use $\mathbf{1}$, $\mathbf{0}$ and $\mathbf{I}$ to denote an all-ones vector, all-zeros vector and an identity matrix with proper sizes, respectively. 
The notation
$Diag(\mathbf{a})$ is used to denote a diagonal matrix with elements of $\mathbf{a}$ on its
diagonal.
The element-wise product and division of $\mathbf{a}$ and $\mathbf{b}$ are denoted by $\mathbf{a} \cdot \mathbf{b}$ and $\mathbf{a} ./\mathbf{b}$, respectively.

\section{SBL and UAMP-SBL with Hyper-parameter Auto-Tuner}

In conventional SBL, the precision vector $\boldsymbol{\gamma}$ is learned by maximizing the logarithm of the a posteriori probability
\begin{equation}\label{newa}
	\begin{aligned}
		\log p(\boldsymbol{\gamma}|\mathbf{y}) &\propto \log p(\mathbf{y}|\boldsymbol{\gamma})+\log p(\boldsymbol{\gamma})
	\end{aligned}
\end{equation}
where the marginal likelihood function is
\begin{equation}
p(\mathbf{y}|\boldsymbol{\gamma})=\int p(\mathbf{y}|\mathbf{x}) p(\mathbf{x}|\boldsymbol{\gamma})d\mathbf{x}.
\end{equation}
As the value of $\boldsymbol{\gamma}$ that maximizes \eqref{newa} cannot be obtained
in closed form, iterative re-estimation is employed, leading to the iterative SBL algorithm shown in Algorithm 1. 
\begin{algorithm}\caption{SBL}\label{SBL}
\textbf{Repeat}
\begin{algorithmic}[1]

\STATE      $\mathbf{Z}=
(\beta\mathbf{A}^H\mathbf{A}+\textit{Diag}(\boldsymbol{\gamma}))^{-1}$\;\label{LINE1}
      
    \STATE     $\hat{\mathbf{x}}=\beta\boldsymbol{Z}\mathbf{A}^H\mathbf{y}$\;\label{LINE2}
     \STATE     $  \gamma_n=(2\epsilon+1)/(2\eta+|\hat{x}_n|^2+Z_{n,n}),n=1,\ldots,N$\;\label{LINE2}   
\end{algorithmic}
\textbf{Until terminated}
\end{algorithm}

To address the high complexity of the SBL algorithm, UAMP-SBL was developed in \cite{luo2021unitary}, leveraging the UAMP algorithm \cite{2015arXiv150404799G}. UAMP-SBL is derived based on a unitary transform of the original model \eqref{sysm}, i.e.,  
\begin{equation}\label{sys_uamp}
	\mathbf{r}  =\mathbf{\Phi}\mathbf{x} + \boldsymbol{\omega}
\end{equation}
where the unitary matrix $\mathbf{U}$ is obtained from the SVD of the measurement matrix
$\mathbf{A} = \mathbf{U \Lambda V}$, $\mathbf{r}=\mathbf{U}^H\mathbf{y}$, $\mathbf{\Phi}=\mathbf{U}^H\mathbf{A}=\mathbf{\Lambda V}$,
$\mathbf{\Lambda}$ is an $M\times N$ rectangular diagonal matrix, and $\boldsymbol{\omega}=\mathbf{U}^H\boldsymbol{\zeta}$ remains a zero-mean Gaussian noise vector with the same covariance matrix $\beta^{-1}\mathbf{I}$.
As
the noise precision $\beta$ is often unknown, its estimation is also considered in developing the UAMP-SBL algorithm \cite{luo2021unitary}. The joint conditional distribution of $\mathbf{x}$,  $\boldsymbol{\gamma}$, $\beta$ and $\mathbf{h}$ can be
expressed as
\begin{equation}
	\begin{aligned}
			p(\mathbf{x},\mathbf{h},\boldsymbol{\gamma},\beta|\mathbf{r})&\propto p(\mathbf{r}|\mathbf{h}, \beta) p(\mathbf{h}|\mathbf{x}) p(\mathbf{x}| \boldsymbol{\gamma})p(\boldsymbol{\gamma}|\epsilon)p(\beta)	\\
	\end{aligned}
 \end{equation}
where 
$\mathbf{h} = \boldsymbol{\Phi} \mathbf{x}$ is an auxiliary variable. 
Employing the structured variational inference (SVI) \cite{jordan1999introduction,winn2005variational}, \cite{xing2012generalized,dauwels2007variational} with the variational distribution
\begin{equation}
\Tilde{q}(\mathbf{x},\mathbf{h},\boldsymbol{\gamma},\beta)=\Tilde{q}(\beta)\Tilde{q}(\mathbf{x},\mathbf{h})\Tilde{q}{(\boldsymbol{\gamma})},
\end{equation}
UAMP-SBL is derived by incorporating UAMP into SVI. 
With SVI, the approximate inference for $\mathbf{x}$ and $\boldsymbol{\gamma}$ is performed alternately, leading to the iterative UAMP-SBL algorithm shown in Algorithm 2. 

In UAMP-SBL, the rate parameter $\eta$ is simply set to 0, and the shape parameter is tuned automatically (as
shown in Line 13) with the empirical rule \eqref{ep}, 
i.e., $	\epsilon$ is tuned iteratively  with a small positive initial value. The auto-tuner also works well for the conventional SBL algorithm, which can be added to Algorithm 1 following Line 3. 


 \begin{algorithm}\caption{UAMP-SBL}\label{Alg}
 	 Unitary transform: $\mathbf{r} = \mathbf{U}^H\mathbf{y}$, where 
   $\mathbf{U}$ is obtained from the SVD $\mathbf{A} = \mathbf{U \Lambda V}$. Define vector $\boldsymbol{\lambda} = \mathbf{\Lambda \Lambda}^H \textbf{1}$. \\
 	 Initialization: $\tau_x^{(0)} = 1$, $\hat{\mathbf{x}}^{(0)}= \textbf{0}$, $\mathbf{s} = \textbf{0}$, $\hat{\beta}=1$, $\hat{\boldsymbol{\gamma}}^{(0)}=\textbf{1}$, $\epsilon^{(0)}=0.001$ and $i = 1$.\\
 	 \textbf{Repeat}
 	\begin{algorithmic}[1]
 		\STATE      $\boldsymbol{\tau}_{p} = \tau_{x}^{(i-1)} \boldsymbol{\lambda}$\;\label{LINE5}
 		
 		\STATE     ${\mathbf{p}} = \mathbf{\Phi} {\hat{\mathbf{x}}}^{(i-1)} - \boldsymbol{\tau}_{p} \cdot \mathbf{s}$\;\label{LINE6}
 		
 		\STATE $\mathbf{v}_h=\boldsymbol{\tau}_p./(\mathbf{1}+\hat{\beta}\boldsymbol{\tau}_p)$\;\label{LINE7}
 		\STATE
 		$\hat{\mathbf{h}}=(\hat{\beta}\boldsymbol{\tau}_p \cdot\mathbf{r} +\mathbf{p})./(\boldsymbol{1}+\hat{\beta}\boldsymbol{\tau}_p)$\; \label{LINE8}
 		\STATE
 		$\hat{\beta}={M}/(||\mathbf{r}-\hat{\mathbf{h}}||^2+\boldsymbol{1}^{H}\mathbf{v}_{h})$\; \label{LINE9}
 		
 		\STATE        ${\boldsymbol{\tau}_s} = \boldsymbol{1}./(\boldsymbol{\tau}_p+\hat{\beta}^{-1}\boldsymbol{1})$\;\label{LINE10}
 		\STATE        ${\mathbf{s}} = \boldsymbol{\tau}_{s} \cdot ( \mathbf{r}-\mathbf{p})$\;\label{LINE11}

 		\STATE        $1/\tau_{q} = (1/N)  \boldsymbol{\lambda}^H \boldsymbol{\tau}_{s}$  \;\label{LINE12}
 		
 		\STATE        ${{\mathbf{q}}} = \hat{\mathbf{x}}^{(i-1)} + \tau_{q} (\mathbf{\Phi}^H \mathbf{s})$ \;\label{LINE13}
 		
 		\STATE $\tau_x^{(i)}= (\tau_{q}/N)\mathbf{1}^{H}(\mathbf{1}./(\mathbf{1}+\tau_q\hat{\boldsymbol{\gamma}}^{(i-1)}))$\;\label{LINE14}
 		 		\STATE $\hat{\mathbf{x}}^{(i)} = \mathbf{q}./(\mathbf{1}+\tau_{q}\hat{\boldsymbol{\gamma}}^{(i-1)})$\;\label{LINE15}
 			 		\STATE $\hat{\gamma}_n^{(i)} = (2\epsilon^{(i-1)}+1)/(|\hat{x}_n^{(i)}|^2+\tau_x^{(i)}),~ n=1,\ldots,N$\;\label{LINE16}
 			\STATE $\epsilon^{(i)} =\frac{1}{2}\sqrt{\log(\frac{1}{N}\sum_n\hat{\gamma}_n^{(i)})-\frac{1}{N}\sum_n\log\hat{\gamma}_n^{(i)}}$\;\label{LINE18}

 		\STATE     $i = i+1$\;\label{LINE30}
 	\end{algorithmic}
\textbf{Until terminated}
 \end{algorithm}


Although the empirical auto-tuner is very effective, which enables the SBL and UAMP-SBL algorithms to deliver outstanding performance, it is interesting to explore potentially more effective auto-tuners, leveraging the powerful NN-based learning technique. 

\section{ Hyper-Parameter Auto-Tuner Learning}

\subsection{NN-Based Auto-Tuner Learning} 

The empirical auto-tuner is significant because it indicates the existence of some function (mapping $\boldsymbol{\gamma}$ to $\epsilon$) that improves the performance of SBL. Inspired by this, we  attempt to find better functions through NN-based learning, i.e., learn a NN-based auto-tuner 
\begin{equation}
\epsilon=\mathcal{NN}( \boldsymbol{\gamma})
\end{equation}
to replace the empirical auto-tuner \eqref{ep} in the UAMP-SBL algorithm. To train the NN, {with the idea of algorithm unrolling \cite{unrolling}},  we unroll the algorithm as shown in Fig. \ref{conv}, which can be regarded as a deep NN.    
Specifically, each
iteration of UAMP-SBL in Algorithm 2 corresponds to one layer of the deep NN. The number of layers (i.e., the number of iterations) is $I$. Each layer consists of a message passing module and a NN module.
\begin{itemize}
    \item The $i$th layer of the deep NN corresponds to the $i$th iteration of UAMP-SBL, i.e., it has inputs $\mathbf{y}$ and $\epsilon^{(i-1)}$, and output $\epsilon^{(i)}$. In the last layer, the estimate of ${\mathbf{x}}$ is the output. 
    \item In the $i$th layer, the message passing module executes the computations (Lines 1 to 12) of the $i$t iteration in Algorithm 2. It is a fixed module, and no parameters need to be learned.
    \item In the $i$th layer, there is a NN module following the message passing module. The NN is the auto-tuner with  {$\hat{\boldsymbol{\gamma}}^{(i)}$ as input and $\epsilon^{(i)}$} as output. The NN needs to be trained. 
    \item As the trained NN-based auto-tuner will be used to replace the empirical auto-tuner in the iterative algorithm, we restrict that the NN modules in all the layers are the same, i.e., they share the same parameters. 
\end{itemize}

\subsection {Architecture for the NN-Based Auto-Tuner}

\begin{figure*}
	\begin{center}
		\includegraphics[width=5.5 in]{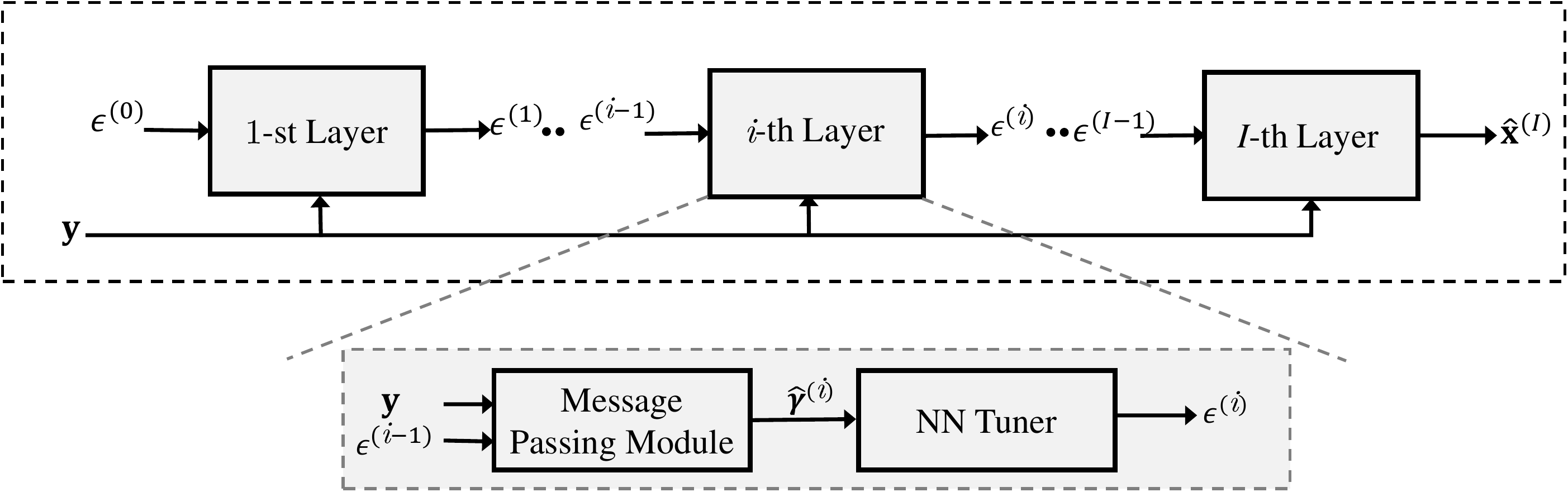}\\
		\caption{{Deep NN architecture obtained through algorithm unrolling.}}\label{conv}
	\end{center}
\end{figure*}

The input to the NN in the $i$th layer is the precision vector $\boldsymbol{\hat\gamma}^{(i)}=[\hat\gamma_1^{(i)},\hat\gamma_2^{(i)},\ldots,\hat\gamma_N^{(i)}]^T$, and the output is {$\epsilon^{(i)}$}. {In Fig. \ref{conv}, the NN has one hidden layer.} 
As we restrict that the auto-tuner does not change over iterations, the NNs in all the layers share the same parameters, so we drop the layer index in this section. The $j$th output in the hidden layer is denoted as    
\begin{equation}\label{z0}
{\psi}_j=g_1(\mathbf{w}_j^T\boldsymbol{\hat\gamma}+{b}_{j}), ~~~~{1 \leqslant j \leqslant L}
\end{equation} 
where $\mathbf{w}_j=[{w}_{j1},{w}_{j2},\ldots,{w}_{jN}]^T$ is a length-$N$ weight vector of the hidden layer, ${b}_{j}$ is the bias at the $j$th neuron, $g_1(.)$ is the  element-wise activation
function of neurons in the hidden layer, and $L$ is the number of neurons of the hidden layer. 
The output layer only has a single hidden neuron, and the output is 
\begin{equation}\label{z0}
	{\epsilon}=g_2(\boldsymbol{\alpha}^T\boldsymbol{\psi}+d),
\end{equation} 
where $\boldsymbol{\psi}=[{\psi}_{1},{\psi}_{2},\ldots,{\psi}_{L}]^T$ is the output of the hidden layer,  $\boldsymbol{\alpha}=[\alpha_1,\alpha_2,\ldots,\alpha_L]^T$ is the input weights of the output layer, $d$ is the bias of the output layer, and $g_2(.)$ is the activation function of the output layer. We use Tanh for the  activation functions $g_1(.)$ and $g_2(.)$ in the NN.  
\begin{figure}
	\begin{center}
		\includegraphics[width=1.95 in]{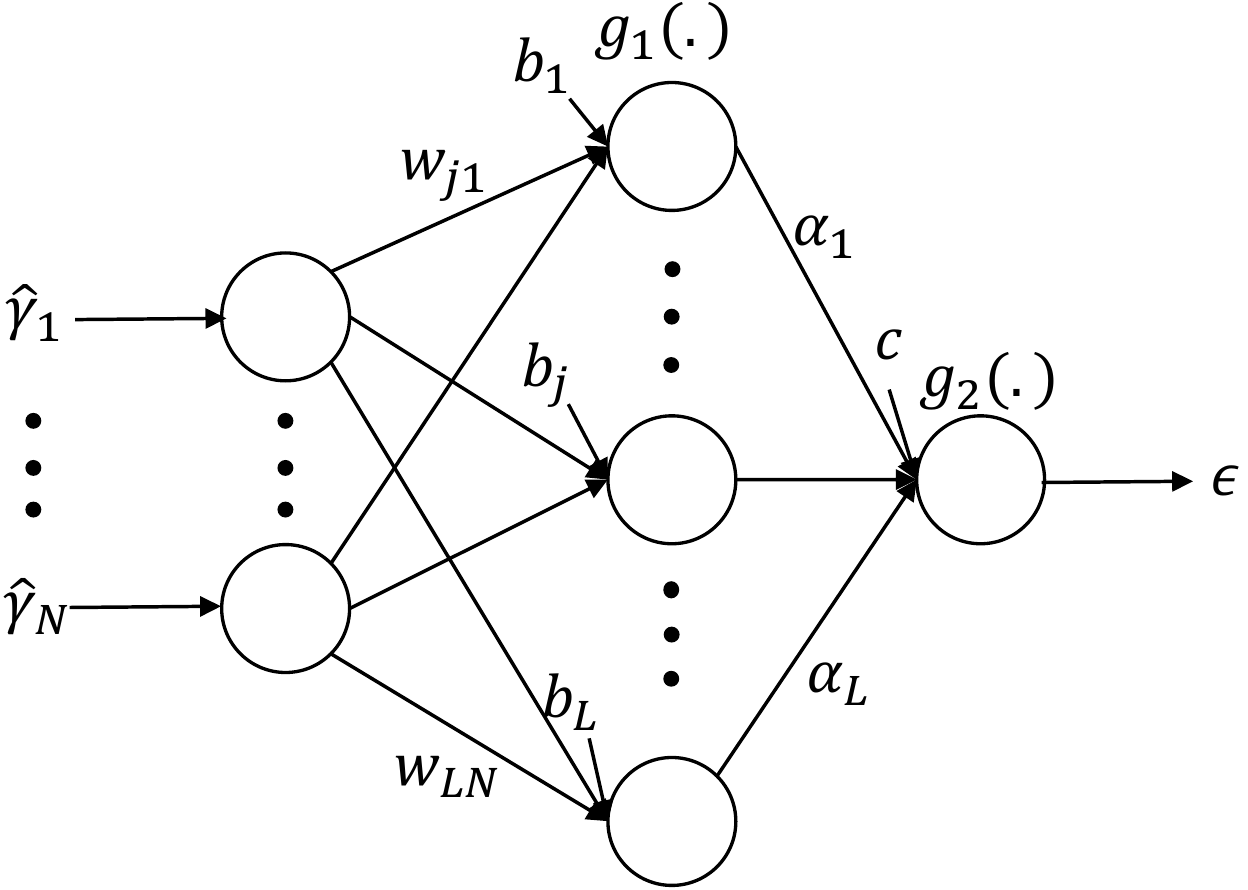}\\
		\caption{Architecture of the NN-based auto-tuner, which is shared by all the layers in Fig. 1 (the layer index is omitted). }\label{snrdB50}
	\end{center}
\end{figure}
\subsection {NN-Based Auto-Tuner Training}
 We use mean square error as the cost function, which is given as
 \begin{equation}
 	\text{Loss}=\frac{1}{\mathcal{M}}\sum^{\mathcal{M}}_{m=1}||\hat{\textbf{x}}(m)-\textbf{x}(m)||^2,
 \end{equation}
where $\hat{\textbf{x}}(m)$ is the output of the deep NN (predicted $\textbf{x}(m)$) for the $m$th training sample, and $\mathcal{M}$ is the number of training samples.     
{The training is performed as follows:}
\begin{itemize}
    \item The training sample pairs $\{\mathbf{x}(m), \mathbf{y}(m)\}$ can be easily generated based on (1) through simulations.
    \item As the empirical auto-tuner, we expect that the trained single NN-based auto-tuner works for different SNRs and sparsity rates. Hence, the training samples are generated with  different SNRs and sparsity rates. This is useful as we may not have the knowledge of sparsity rate and SNR in practical applications. 
    \item As mentioned before, the NNs in all layers share the parameters {(i.e., $\{\mathbf{w}_j,{b}_j,\boldsymbol{\alpha},d\})$}, so we tie the parameters in the training process with back-propagation. The number of parameters to be trained in the whole network is $\mathcal{O}(NL)$ (instead of $\mathcal{O}(INL)$), making the training process more efficient. 
\end{itemize}




\section{Numerical Results }

We compare the NN-based auto-tuner and the empirical auto-tuner for SBL and UAMP-SBL using normalized mean squared error (NMSE), which is defined as
\begin{equation}
	\text{NMSE}=\frac{1}{\mathcal{N}}\sum^{\mathcal{N}}_{n=1}||\hat{\mathbf{x}}(n)-\mathbf{x}(n)||^2/||\mathbf{x}(n)||^2,
\end{equation}
where $\mathcal{N}$ is the number of test samples. 
As a performance benchmark, the support-oracle bound 
is also included. We set $M=80$ and $N=100$. 
The number of hidden nodes is 256.
The vector $\mathbf{x}$ is drawn from
a Bernoulli-Gaussian distribution with a non-zero probability $\rho$. For both SBL and UAMP-SBL, we set the number of iterations {$I=50$}. The SNR is defined as $\text{SNR}= E||\mathbf{Ax}||/E||\boldsymbol{\zeta}||^2$.




The deep learning framework \textit{Tensorflow} is used for training. Batch gradient descent is employed, and cross-validation is used to avoid overfitting. The dataset is generated with combinations of  $\text{SNR}=\{10,20,30,40,50\}$ and sparsity rate $\rho=\{0.1,0.2,0.3,0.4,0.5\}$, and the size of the dataset is 50000. We use 40\%, 40\% and 20\% of the dataset for training, validation and testing, respectively. Through the validation data set, we determine that the batch size is 32 and the times of epoch is 150.  Adam optimizer with a learning rate 0.01 is employed to update parameters.

We first assume that the measurement matrix $\mathbf{A}$ is correlated, which is constructed using $\mathbf{A}=\mathbf{C}_L^{1/2}\mathbf{G}\mathbf{C}_R^{1/2}$, where $\mathbf{G}$ is an i.i.d. Gaussian matrix with mean zero and unit variance, and $\mathbf{C}_L$ is an $M\times M$ matrix with the $(m,n)$th element given by $c^{|m-n|}$, where $c$ here is set as 0.1. 
Matrix $\mathbf{C}_R$ is generated in the same way
but with a size of $N\times N$.  Fig. 3 shows the performance of UAMP-SBL with the empirical and the learned NN-based auto-tuners, where $\rho=0.1$. It can be seen that, at $\text{SNR} =50\text{dB}$, UAMP-SBL with both auto-tuners approaches the bound, but the trained auto-tuner converges faster. At $\text{SNR}=  15\text{dB}$, UAMP-SBL with the trained auto-tuner considerably outperforms the algorithm with the empirical auto-tuner. With correlated $\mathbf{A}$, we also examine the performance of the conventional SBL algorithm with both auto-tuners and fixed shape parameter $\epsilon=10^{-4}$. The results are shown in Fig. \ref{Fig.sub.41} and Fig. \ref{Fig.sub.42} with $\text{SNR} =50\text{dB}$ and $\text{SNR} =15\text{dB}$, respectively. It can be seen that the SBL with both trained auto-tuners delivers significantly better performance than the SBL with fixed $\epsilon$. At $\text{SNR} =15\text{dB}$, the SBL with learned auto-tuner performs better than that with the empirical auto-tuner.   

\begin{figure}[H]
\centering  
\subfigure[SNR=50dB]{
\label{Fig.sub.31}
\includegraphics[width=0.24\textwidth]{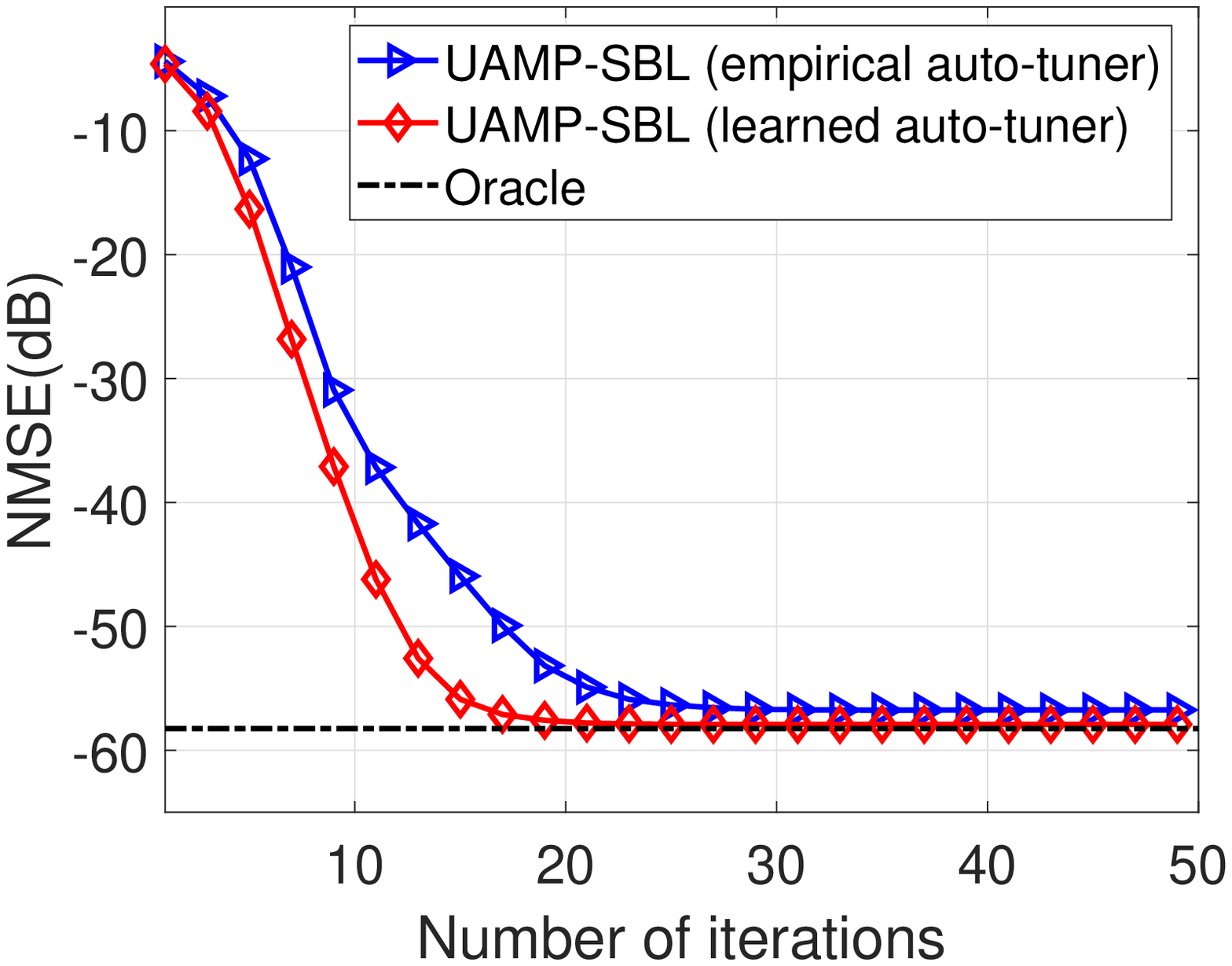}}
\hspace{-10pt}
\subfigure[SNR=15dB]{
\label{Fig.sub.32}
\includegraphics[width=0.24\textwidth]{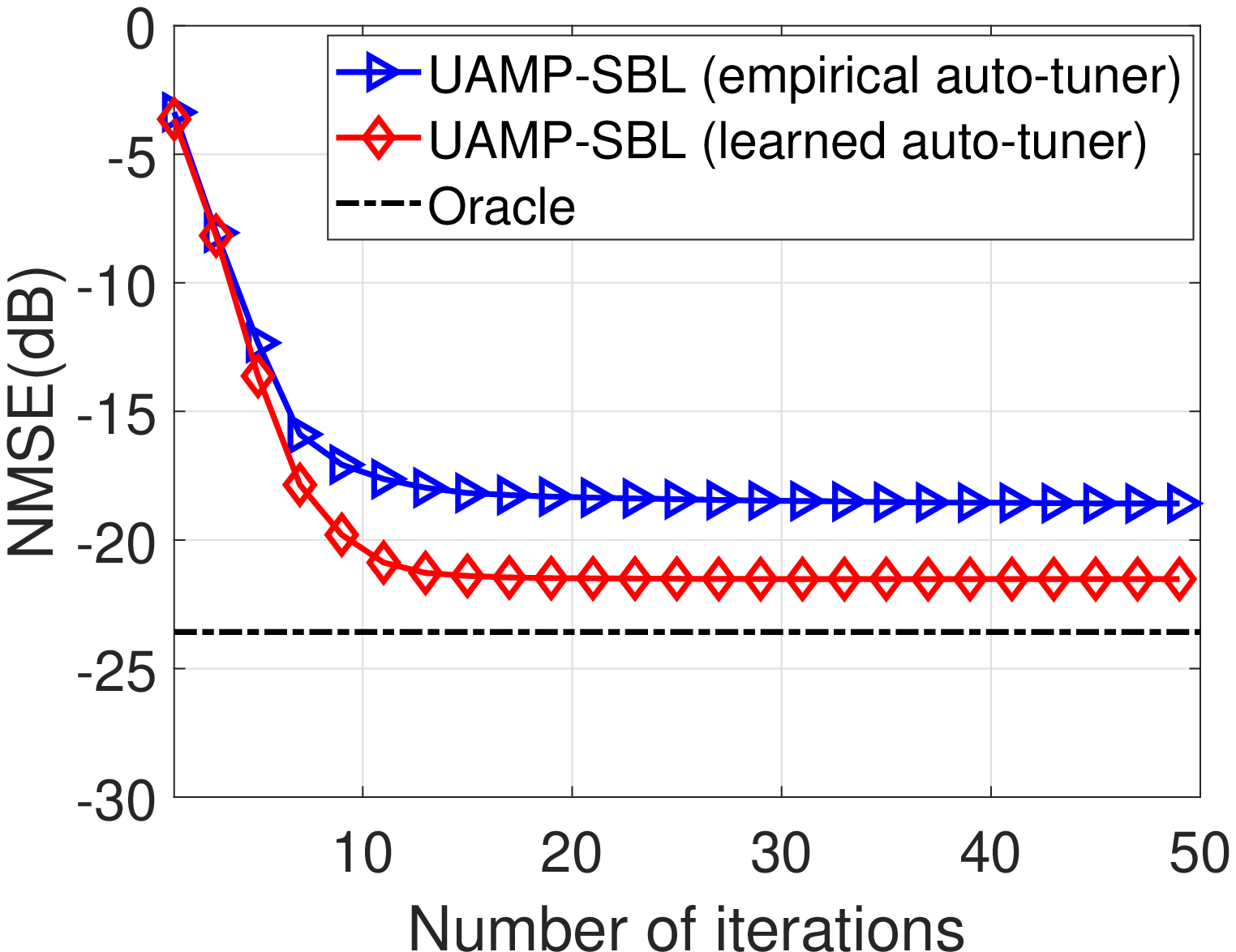}}
\caption{NMSE performance of UAMP-SBL with the empirical auto-tuner and the learned auto-tuner at SNR=50dB and 15dB, respectively.}
\label{Fig.main3}
\end{figure}
\begin{figure}[H]
\centering  
\subfigure[SNR=50dB]{
\label{Fig.sub.41}
\includegraphics[width=0.24\textwidth]{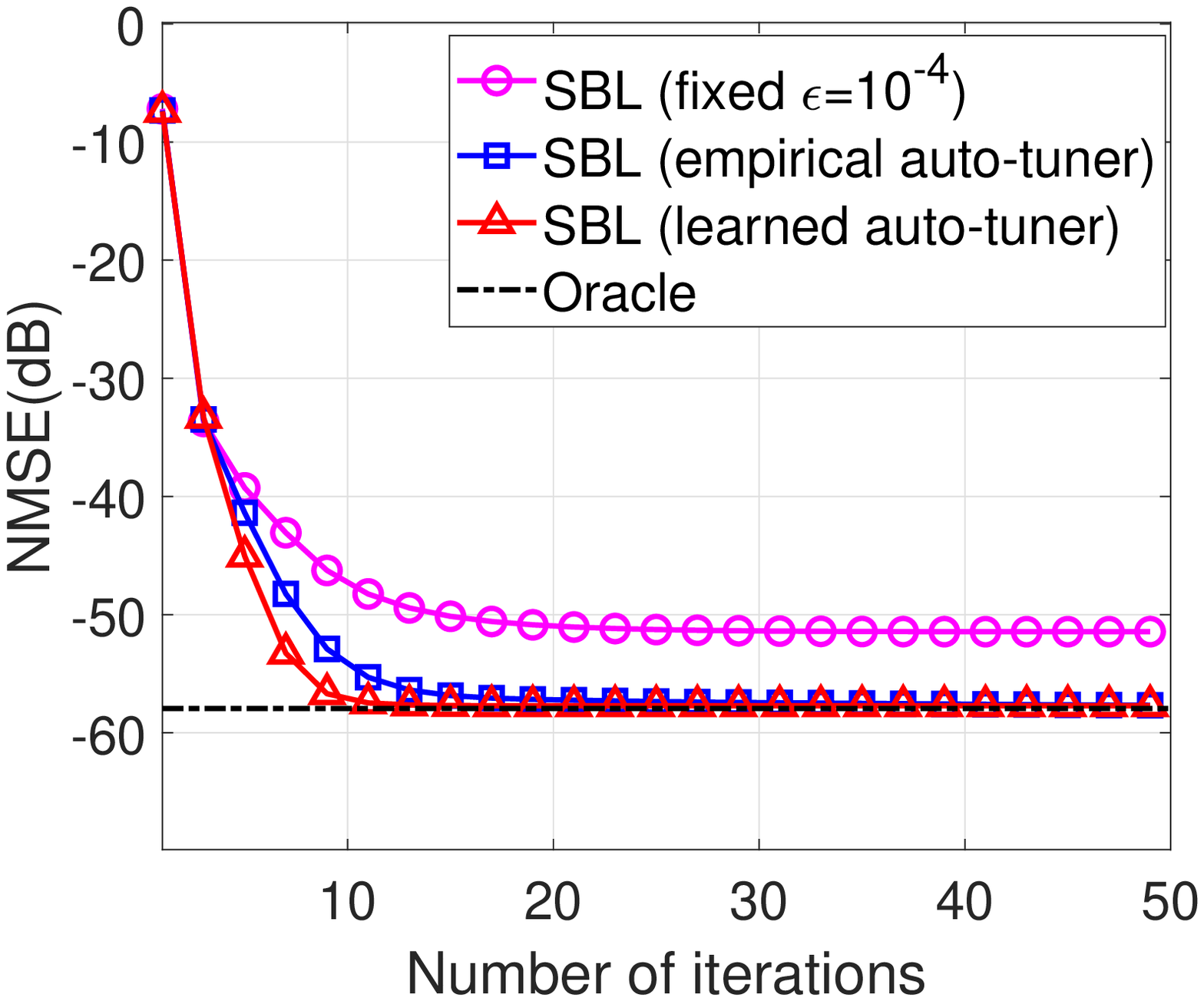}}
\hspace{-10pt}
\subfigure[SNR=15dB]{
\label{Fig.sub.42}
\includegraphics[width=0.24\textwidth]{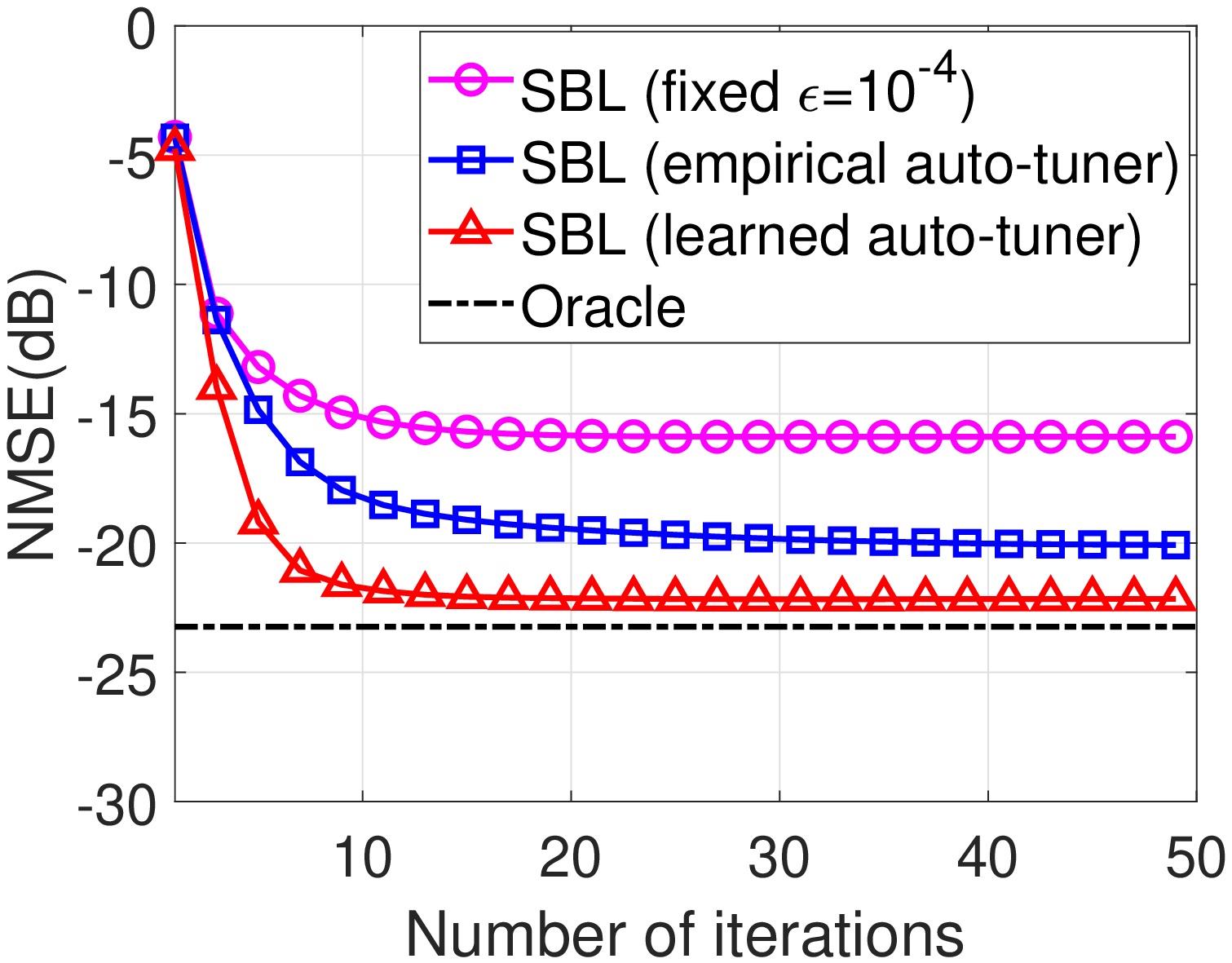}}
\caption{NMSE performance of SBL with the empirical auto-tuner and the learned auto-tuner at SNR=50dB and 15dB, respectively.}
\label{Fig.main4}
\end{figure}

\begin{figure}
	\begin{center}
		\includegraphics[width=3.1in]{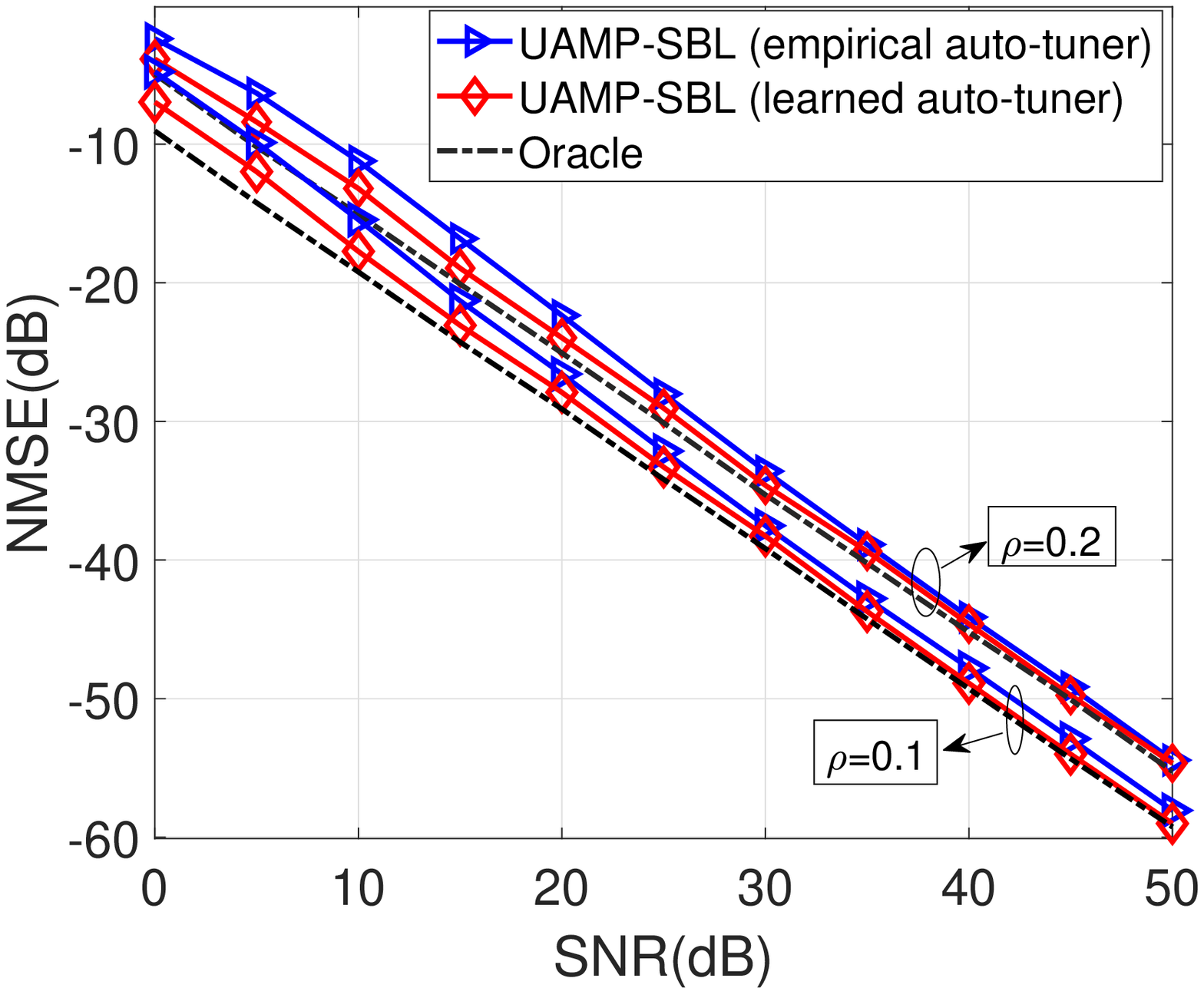}\\
		\caption{NMSE performance of UAMP-SBL with the empirical auto-tuner and trained auto-tuner with different SNRs.}\label{UAMPvsSNR_ratio}
	\end{center}
\end{figure}	
\begin{figure}
	\begin{center}
		\includegraphics[width=3.1in]{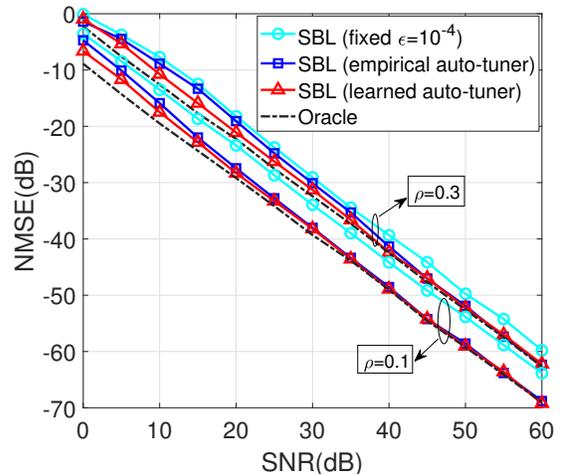}\\
		\caption{NMSE performance of SBL with the empirical auto-tuner and the trained auto-tuner with different SNRs.}\label{SBL_vsSNR_ratio}
	\end{center}
\end{figure}

Figures \ref{UAMPvsSNR_ratio} and \ref{SBL_vsSNR_ratio} show the performance of UAMP-SBL and the conventional SBL versus SNR, where the elements of the measurement matrix are independently drawn from the standard Gaussian distribution. In Fig. \ref{UAMPvsSNR_ratio}, $\rho=0.1$ and 0.2 are considered. It can be seen that UAMP-SBL with learned auto-tuner outperforms the one with the empirical auto-tuner in the low SNR region. Fig. \ref{SBL_vsSNR_ratio} includes the performance of the conventional SBL with fixed shape parameter $\epsilon=10^{-4}$, and two auto-tuners, where $\rho=0.1$ and 0.3 are considered.  Again, we can see that the SBL with auto-tuners significantly outperforms that with fixed shape parameter. In addition, the SBL with learned auto-tuner still works better than that with the empirical one in low SNR range.  



\section{Conclusion}
Inspired by the empirical auto-tuner in \cite{luo2021unitary}, we developed a NN-based learning technique for hyper-parameter auto-tuning in SBL. {Our simulations show that the empirical auto-tuner in \cite{luo2021unitary} works fairly well, and is attractive due to its simple closed form. However, }through training, we can find better hyper-parameter auto-tuners, leading to considerably improved convergence speed or performance in the low SNR region.


\ifCLASSOPTIONcaptionsoff
  \newpage
\fi
\normalem
\bibliographystyle{IEEEtran}
\bibliography{IEEEabrv,bare_jrnl}
\end{document}